# New limits on WIMP interactions from the SIMPLE dark matter search

**TA Girard[1], T. A. Morlat**
*Center for Nuclear Physics & Department of Physics, University of Lisbon*
*Av. Prof. Gama Pinto 2, 1649-003 Lisbon Portugal*
*E-mail:* criodets@cii.fc.ul.pt

**M. Felizardo**
*Department of Physics, New University of Lisbon, 2829-516 Caparica, Portugal*
*E-mail:* felizardo@itn.pt

**A. C. Fernandes, A. R. Ramos, J. G. Marques**
*Instituto Tecnologico e Nuclear and Center for Nuclear Physics, University of Lisboa*
*Estrada Nacional 10, 2686-953 Sacavém, Portugal*
*E-mail:* anafer@itn.pt, arial@itn.pt, jmarques@itn.pt

**(for the SIMPLE Collaboration)**

The SIMPLE project uses superheated $C_2ClF_5$ liquid detectors to search for particle dark matter candidates. We report the results of the first stage exposure (14.1 kgd) of its latest two-stage, Phase II run, with 15 superheated droplet detectors of total active mass 0.208 kg. In combination with the results of the neutron-spin sensitive XENON10 experiment, these results yield a limit of $|a_p| < 0.32$, $|a_n| < 0.17$ for $M_W = 50$ GeV/c$^2$ on the model-independent, spin-dependent sector of weakly interacting massive particle (WIMP)-nucleus interactions, and together yield a 50% reduction in the previously allowed region of the phase space. The result provides a contour minimum of $\sigma_p \sim 2.8 \times 10^{-2}$ pb at $M_W = 45$ GeV/c$^2$, constituting the most restrictive direct detection limit to date against a spin-dependent WIMP-proton coupling. In the spin-independent sector, the result is seen to offer the prospect of contributing to the question of light mass WIMPs with an improvement in the current understanding of its nucleation efficiency.

---

[1] Speaker

## 1. Introduction

We report the first stage results of a two stage SIMPLE measurement obtained from a total exposure of 14.10±0.01 kgd with 15 superheated droplet detectors (SDDs) containing a total active mass of 0.208 kg $C_2ClF_5$.

The SIMPLE [1,2] detector, in contrast to PICASSO [3] and COUPP [4], is based on a 1-2% suspension of superheated liquid $C_2ClF_5$ droplets (<r>~30 μm radius) in a viscoelastic 900 ml gel matrix, which undergo transitions to the gas phase upon energy deposition by incident radiation. The physics of superheated liquids is however generic: nucleation of the gas phase in the liquid requires [5]: (i) that the energy deposited be greater than a thermodynamically-defined minimum energy, and (ii) that this energy be deposited within a thermodynamically-defined minimum distance inside the droplet. Together, energy depositions of order ~ 150 keV/μm are required for a bubble nucleation, which renders the technique effectively insensitive to the majority of traditional light particle backgrounds which complicate more conventional dark matter search detectors (including electrons, γ's and cosmic muons). The insensitivity is not trivial, equaling an intrinsic rejection factor superior to that of other search techniques by 1-5 orders of magnitude.

## 2. Experimental Detail

The SDDs were operated in the 1500 mwe GESA facility of the Laboratoire Souterrain à Bas Bruit [6] in southern France, the layout of which is shown in Fig. 1. The cavern is shielded from the rock environment by a 30-100 cm thickness of concrete, which is internally sheathed by a 1 cm thickness of iron. The SDDs were immersed to a depth of 20 cm in a temperature-controlling, 700 liter water pool within the cavern, which rested on a dual vibration absorber placed atop a 20 cm thick wood platform resting on a 50 cm thick concrete floor. The pool was surrounded by layers of sound and thermal insulation. An additional 50-75 cm thick water shielding surrounded the pool and platform, with a 75 cm water thickness overhead; 50 cm of water separated the pool bottom from the detector bases, as shown in the schematic. Monte-Carlo simulations of the on-detector neutron field [7], based on radioassays of the shielding materials and which account for spontaneous fission, decay-induced (α,n) reactions and (μ,n) reactions in the rock, show negligible variations for concrete thicknesses ≥ 20 cm, and yielded an expected neutron background of 1.09 ± 0.02 (stat) ± 0.07 (syst) evt/kgd.

The ambient radon level varies seasonally between 28–1000 Bq/m$^3$ as a result of water circulation in the mountain. To reduce this, the cavern air was vented at ~ 0.2 m$^2$/s, reducing the ambient radon levels to 50-80 Bq/m$^3$ during the measurement. Diffusion of the environmental radon into a detector is limited by the surrounding waterpool, which covered the detectors to 4 cm above their glycerin levels, and which was circulated at 25 liter/min (equivalent to replacing the top 1 cm water layer each minute) through the temperature-controlling external cryothermostat, preventing equilibrium buildup. The radon contribution is further reduced by the short radon diffusion lengths of the SDD construction materials (glass, plastic, metal), the $N_2$ overpressuring, and the glycerin layer covering the gel. Combined, the overall $\alpha$ contribution to the measurement, including the radon progeny and detector contribution, is estimated at 3.26 ± 0.08 (stat) ± 0.76 (syst) evt/kgd.

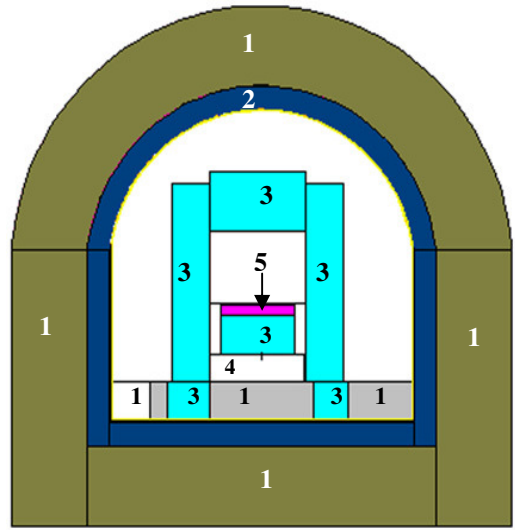

**Fig. 1.** Schematic of the SIMPLE SDDS in the GESA site: 1-concrete, 2-steel sheath, 3-water, 4-wood platform, 5-SDDS.

Each SDD was fabricated in an underground (210 mwe) clean room near GESA, capped with a new MCE-200 electret microphone [8] and immersed to a depth of 20 cm in the GESA water pool; each was pressurized to 2 bar and maintained at 9.0±0.1°C by the water pool to provide a recoil threshold energy of 8 keV. The fabrication procedures have been previously described elsewhere [1]; the gel ingredients, all biologically-clean food products, were purified using actinide-specific ion-exchanging resins. The freon was single distilled; the water, double distilled. The presence of U/Th contaminations in the gel, measured at ~ 0.1 ppb by low-level $\alpha$ and $\gamma$ spectroscopy of the production gel, yielded an overall $\alpha$-background level of < 0.5 evt/kg/d. A similar level is measured for the detector containment materials.

## 3. Analysis

Data taken between 27 October 2009 and 5 February 2010 were analyzed for this report [9]. Data losses during the period, resulting from the detectors being introduced at one device per day over the three week installation period, and from weather-induced power failures, provide a net exposure of 14.10±0.01 kgd.

The SDD signals, pressures and temperature were monitored continuously during operation, as also the radon level. Each detector was first inspected for raw signal rate and pressure evolution over the measurement period, and an initial data set (4056 events) extracted. These were analyzed according to their signal characteristics for origin as described elsewhere in these Proceedings [9], yielding 60 particle-induced events.

The signal events were compared with similarly-analyzed neutron and α calibration signals [9], obtained from weak sources of Am/Be and $U_3O_8$ respectively, which yielded 14 events most likely of neutron origin, or an efficiency-corrected 0.99 ± 0.27 (stat) evt/kgd, with an acceptance of ≥ 97%. At 9°C, the reduced superheat of the devices is 0.3, and the probability of events from electrons, γ's and mip's negligible [1] over the exposure.

An upper limit on the number of WIMP events is estimated by applying the Feldman-Cousins method [10], based on the observation of 14 events with a background one standard deviation below the central value of the expected neutron background. This gives 4.3 events, and a WIMP rate of 0.57 evt/kgd at 90% C.L.

## 4. Results and Conclusion

The impact of the result in the spin-dependent (SD) phase space is shown in Fig. 2 for $M_W$ = 50 GeV/$c^2$, calculated using a model-independent formulation [11] in which the region excluded by an experiment lies outside the indicated area, and the allowed region is defined by the intersection of the various areas. $M_W$ above or below this choice yield less restrictive limits [11].

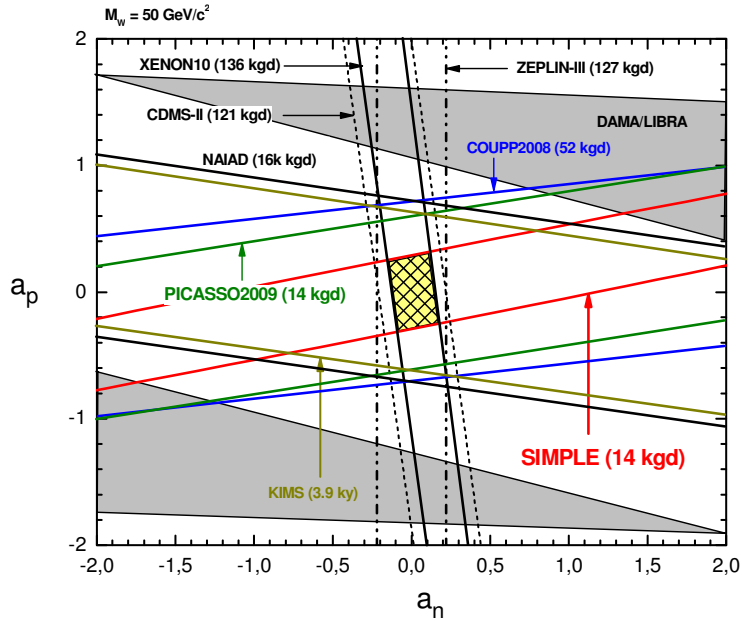

**Fig. 2:** SD $a_p$-$a_n$ for SIMPLE at $M_W$ = 50 GeV/$c^2$, together with benchmark experiment results. The allowed regions are defined by a band (single nuclei target) or ellipse (multinuclei target), with the external area excluded. The cross-hatched central area about (0,0) indicates the region allowed by this result and XENON10.

The calculations employ a standard isothermal halo, nucleation efficiency $\eta = 1-E_{thr}/E$ [1], and WIMP scattering rate [12] with zero momentum transfer spin-dependent cross section for elastic scattering:

$$\sigma_{SD} \sim [a_p\langle S_p\rangle + a_n\langle S_n\rangle]^2 (J+1)J^{-1}, \qquad (1)$$

where $E_{thr}$ is the threshold recoil energy, $a_{p,n}$ are the WIMP-proton,neutron coupling strengths, $<S_{p,n}>$ are the expectation values of the proton (neutron) group's spin, and J is the total nuclear spin. The form factors of Ref. [12] have been used for all odd-A nuclei. The spin values of Strottman have been used for $^{19}$F [13]; for $^{35}$Cl and $^{37}$Cl, Ref. [11], while for $^{13}$C these were estimated by using the odd group approximation. The shaded area represents the allowed DAMA/LIBRA region [14]. As indicated, the present result combined with XENON10 [15] yields limits of $|a_p| \leq 0.32$, $|a_n| \leq 0.16$ on the SD sector of WIMP-nucleus interactions for $M_W = 50$ GeV/c$^2$, with ~ 50% reduction in the allowed region of the phase space.

For comparison with the recent reportings by other experiments, we show in Fig. 3 the SD WIMP-proton scattering cross section as a function of WIMP mass. Although this representation assigns the full interaction strength to the proton [11], neglecting entirely any contribution from the almost equal neutron spin presence, the result is clearly seen to provide the most restrictive direct detection constraint on the cross section; it moreover begins to converge with the realm of indirect search results such as SuperK [16] and IceCube [17].

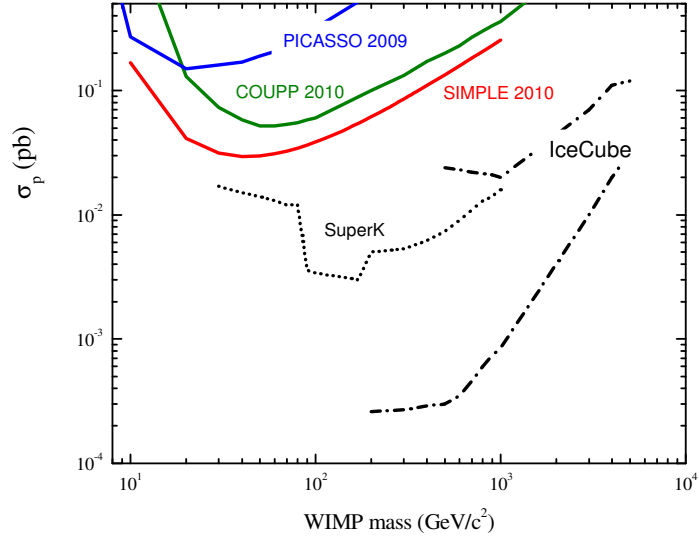

**Fig. 3:** $\sigma_p$ for SIMPLE from this report, together with benchmark experiment results [3,18], including indirect search activities; the allowed regions lie below the respective contours.

As seen in Fig. 4, the impact of the SIMPLE contour in the spin-independent (SI) sector, calculated following the procedures of Ref. [12], is lessened by its light nuclei targets, and by its comparatively small exposure. As evident, the sensitivity at low $M_W$ is weak despite the measurement $E_{thr} = 8$ keV. This is because of η, a theoretical approximation which tends to underestimate the efficiency [19] in this region; the efficiency has more recently been shown [20,21] to behave as $\eta' \sim 1-e^{\delta\eta}$, where δ describes the statistical nature of the deposited energy and heat conversion, and is a detector-specific response parameter. As illustrated by the dotted contour calculated with η=1, an improved determination of η is tantalizingly seen to offer an additional contribution to the question of light mass candidates recently generated by the concordance between CDMS, CoGeNT and DAMA/LIBRA [22,23].

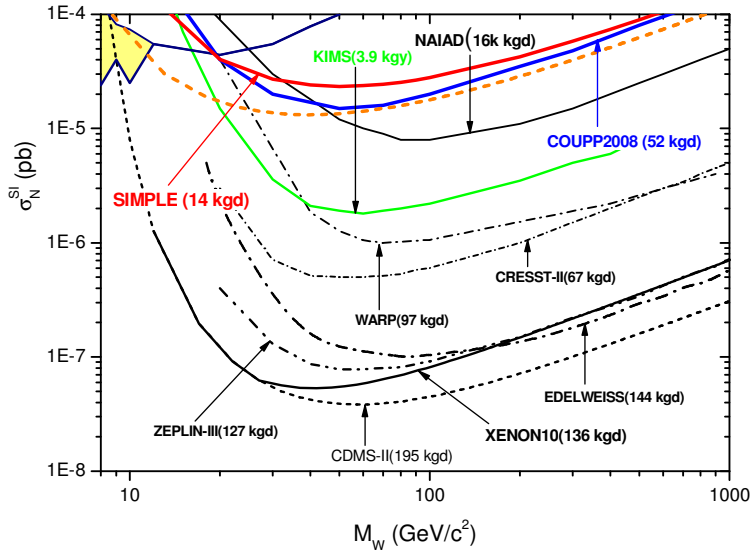

**Fig. 4:** SI limit contour for SIMPLE; the lower dashed contour represents the current result with η=1. Also shown are several of the leading SI search results, with the shaded area representing the recent result of CoGeNT [24].

In summary, a conservative analysis of the first 14.1 kgd of data from Phase II SIMPLE yields the most restrictive limits to date on the WIMP-proton parameter space of SD WIMP interactions, and further demonstrate the competitiveness of the superheated liquid technique in the search for astroparticle dark matter in both SD and SI sectors.

The above results are constrained by the current neutron background level in the measurement and by η. Previous neutron calibration studies are being reexamined to determine the nucleation efficiency for the SIMPLE detectors, which is anticipated to yield an automatic improvement in both these and the recently-completed Stage 2 results. Reduction of the neutron background has been effected in the Stage 2 measurements via an increase in the shielding beneath the waterpool, estimated to reduce the overall neutron contribution by ~ factor 5; a preliminary analysis of the Stage 2 data identifies 3 observed low amplitude events of neutron origin, consistent with the MCNP projections.